\documentclass[12pt]{article}

\usepackage{amssymb}
\usepackage{color,graphicx}
\usepackage{amsmath}

\usepackage{hyperref}

\renewcommand{\thefootnote}{\fnsymbol{footnote}}

\begin{document}

\renewcommand{\abstractname}{\hfill}

%%%% *****************************************************************
%%%% *************    Text stat'i     ********************************
%%%% *****************************************************************
\newpage
\pagenumbering{arabic}

\begin{center}
\bf
A. A. Grib\,\footnote{\,Herzen Russian State Pedagogical University,
St. Petersburg, Russia, e-mail:\, andrei\_grib@mail.ru}\footnote{A.A.\,Friedmann
Laboratory of Theoretical Physics, St. Petersburg, Russia.},
Yu. V. Pavlov\,\footnote{\,Institute for Problems of Mechanical Engineering,
Russian Academy of Sciences, St. Petersburg, Russia,
e-mail:\, yuri.pavlov@mail.ru}
\footnote{\,Lobachevskii Institute of Mathematics and Mechanics,
Kazan Federal University, Kazan, Russia.}
\end{center}

\begin{center}
{\Large \bf {Can the energy of a particle be negative\\[4pt]
in the absence of external
fields?}}\,\footnote{\href{https://doi.org/10.1134/S0040577923090088}{{\it
Theoretical and Mathematical Physics}, Vol.~{\bf 216}(3): 1337--1348 (2023)}
\ (Translated from Russian in
\href{https://doi.org/10.4213/tmf10453}{{\it Teoreticheskaya i
Matematicheskaya Fizika}, Vol.~{\bf 216}, No. 3, pp. 504--518, 2023)}.}
%% Received January 29, 2023. Revised February 24, 2023. Accepted March 7, 2023.
\end{center}

\begin{abstract}
    We study the question of conditions for the existence of negative-energy
states of particles in the absence of external fields in inertial and
noninertial frames of reference.
    We show that in the nonrelativistic case in noninertial reference frames,
there always exist domains where the energy of particles is negative.
    We also show that in the relativistic case, the existence of
negative-energy states of point particles does not lead to violations of
the energy dominance condition.
    We consider conditions for the appearance of negative and zero energies
of particles in the Milne universe and Rindler space-time.
\end{abstract}

{\small
\\

DOI:\, \href{https://doi.org/10.1134/S0040577923090088}{10.1134/S0040577923090088}
}

%%%% *****************************************************************
\section{\normalsize Introduction}

\hspace{\parindent}
    The sign of a particle energy is important from the standpoint of
describing the properties of the particle motion and possible physical
phenomena in a reference frame under consideration.
    If the particle energy is assumed to be zero at an infinite distance
from force centers, then negative-energy states in a force field correspond
to bounded trajectories~\cite{LL_I}.
    Positive energies correspond to unbounded motion.
    In the relativistic case, the presence of negative-energy states leads
to the possibility of extracting energy from a rotating black hole
(the Penrose effect~\cite{Penrose69}).
    A similar effect occurs in the case of charged black
holes~\cite{ChristodoulouRuffini71}.
    In quantum field theory, the presence of negative-energy states leads
to particle production effects in the external field~\cite{GMM}.

    In a noninertial reference frame, the presence of negative-energy states
causes new physical effects, for example, the Unruh effect~\cite{Unruh76}.
    The authors of~\cite{GribPavlov2016d} showed that in a uniformly rotating
coordinate system in flat space, as well as in the Kerr black hole metric,
there is a surface (static limit) beyond which no body can be at rest and,
as in the ergosphere of a black hole~\cite{MTW},
negative-energy particle states are possible.
    For a rotating observer, this can lead to phenomena that are similar to
the Penrose effect~\cite{GribPavlov2019}.

    There are still no experimental observations of the Penrose or Unruh
effects, nor of the widely known Hawking effect of black hole
radiation~\cite{Hawking75}.
    However, most physicists do not doubt their existence.
    In this paper, we consider the question of whether negative energy
values can occur in inertial reference frames and analyze several examples
where negative energies appear in noninertial frames in the nonrelativistic
and relativistic cases.

\vspace{4mm}
%%%% *****************************************************************
\section{\normalsize Energy of a particle in an inertial reference frame}

\hspace{\parindent}
    In an inertial frame of reference, the Lagrange function of
a nonrelativistic particle of mass $m$ is given by~\cite{LL_I}
    \begin{equation}    \label{nm1}
L = \frac{mv^2}{2} - U,
\end{equation}
    where $\mathbf{v}$ is the particle velocity and
    $U$ is its potential energy in the external force field.
    The particle energy
    \begin{equation}    \label{nm2}
E = \mathbf{v} \frac{\partial L}{\partial \mathbf{v}} - L =
\frac{mv^2}{2} + U
\end{equation}
    is the sum of the kinetic and potential energies, and in the absence of
a force field, it is obvious that it cannot be negative for a positive value
of its mass.

     If the external field corresponds to attraction, then the potential
energy is negative; in this case, the total energy of the particle can also be
negative and unbounded from below if the potential energy $U$ is
unbounded from below.
    For example, in the external field of an attracting body of mass~$M$,
    \begin{equation}    \label{nm6}
E = \frac{mv^2}{2} - G \frac{mM}{r} < - mc^2  \ \ \Rightarrow \ \
r< \frac{GM}{c^2} = \frac{r_g}{2},
\end{equation}
    where $G$ is the gravitational constant, $c$ is the speed of light, and
$r_g$ is the gravitational radius.
    If the potential energy corresponds to the attraction of two elementary
point charges, then
    \begin{equation}    \label{nm7}
E = \frac{mv^2}{2} - \frac{1}{4 \pi \varepsilon_0} \frac{e^2}{r} < - mc^2  \ \
\Rightarrow \ \ r< \frac{e^2}{4 \pi \varepsilon_0 m c^2} ,
\end{equation}
    where $\varepsilon_0$ is the electrical constant and
$e$ is the elementary electric charge.
    We can see from these examples that negative energies comparable in
absolute value to the particle rest energy $mc^2$ appear at the distances
at which the nonrelativistic approximations cannot be used, but the effects
of general relativity and quantum theory must be taken into account.

    The action for a free relativistic particle is given by~\cite{LL_II}
    \begin{equation}    \label{rm01}
S = - m c \int \limits_a^b d s,
\end{equation}
where the integral is taken along the world line between two fixed events
world points) $a$ and $b$, i.e., the initial and final points at which
the particle occurs at certain times $t_1$ and $t_2$.
    With such a choice of the action, the Lagrange function for the free
relativistic particle becomes
    \begin{equation}    \label{rm02}
L =- m c^2 \sqrt{ 1 - \frac{v^2}{c^2}},
\end{equation}
    and the energy
    \begin{equation}    \label{rm03}
E = \mathbf{v} \frac{\partial L}{\partial \mathbf{v}} - L =
\frac{m c^2}{ \sqrt{ 1 - \frac{v^2}{c^2}} } .
\end{equation}
    takes positive values for a positive mass~$m$.

    By definition, the momentum vector is
    \begin{equation}    \label{rm04}
\mathbf{p} = \frac{\partial L}{\partial \mathbf{v}} =
\frac{m \mathbf{v}}{ \sqrt{ 1 - \frac{v^2}{c^2}} } .
\end{equation}
    Energy~(\ref{rm03}) and momentum~(\ref{rm04}) are components of
the energy-momentum 4-vector $(p^i) = (E/c,\mathbf{p} ) $, $i=0,1,2,3$.
    We note that for any proper Lorentz transformation, the sign of the
particle energy is conserved for a particle moving slower than the speed
of light.

    In theoretical physics, one also considers hypothetical particles moving
faster than light (see~\cite{EinstainSb} and the references therein).
    They are called tachyons~\cite{Feinberg67}.
    The connection of tachyons with instability of physical systems was
discussed in~\cite{Kirzhnits96}.
    As is known, the group speed of waves for fields with a negative mass
squared term exceeds the speed of light.
    The fields with such properties are used in the modern standard model of
elementary particles involving the Higgs symmetry breaking
mechanism~\cite{BogoliubovShQF}.
    Under symmetry restoration at high energies, the Higgs field can exhibit
specific properties associated with a negative value of the squared mass term.
    As we showed in~\cite{GribPavlov2022}, the phase transitions with
the restoration of the symmetry of strong (quark-gluon) interactions or
electroweak interactions are possible in particle collisions near astrophysical
black holes, and in that case, the effects associated with a negative squared
mass of the corresponding field can be available for observation.

    The energy-momentum vector of tachyons has the form
    \begin{equation}    \label{rm05}
(p^i)  = \left( \frac{E}{c}, \mathbf{p} \right) =
\frac{m }{ \sqrt{ \frac{u^2}{c^2} - 1 }} \left( c, \mathbf{u} \right) ,
\end{equation}
    where $m$ is a constant (the imaginary part of the tachyon mass) and
$\mathbf{u}$ is the velocity of the tachyon ($u>c$).
    The components of the tachyon energy-momentum satisfy the relation
    \begin{equation}    \label{rm06}
E^2 - \mathbf{p}^2 c^2 = - m^2 c^4.
\end{equation}
    As was first noted in~\cite{Tolman1917}, in passing to another inertial
reference frame moving with some speed~$V$, less than the speed of light,
the order of the time sequence of events varies along the tachyon trajectory.
    If in the initial reference frame $K$ the tachyon was at a point $x_a$
at a time $t_a$ and at a point $x_b$ at a time $ t_b > t_a $,
then in the reference frame $K'$ moving in the same direction as the tachyon,
we obtain
    \begin{equation}    \label{rm07}
\Delta t' = t_b' -t_a' =
\frac{t_b - \frac{V}{c^2} x_b}{\sqrt{1 - \frac{V^2}{c^2}}} -
\frac{t_a - \frac{V}{c^2} x_a}{\sqrt{1 - \frac{V^2}{c^2}}}
= \frac{1 - \frac{uV}{c^2}}{\sqrt{1 - \frac{V^2}{c^2}}} \Delta t,
\end{equation}
    where $u$ is the speed of the tachyon in the reference frame $K$.
    If $V > c^2/u $, then $\Delta t' < 0$ and $t_b' < t_a'$,
reversing the order of events in the original fixed reference frame~$K$.
    Under the transformation to the moving reference frame~$K'$, we obtain
    \begin{equation}    \label{rm08}
E' = \frac{E - Vp}{\sqrt{1 - \frac{V^2}{c^2}}}, \ \ \
p' = \frac{p - \frac{V}{c^2} E}{\sqrt{1 - \frac{V^2}{c^2}}}, \ \ \
u' = \frac{u - V}{1 - \frac{u V}{c^2}}.
\end{equation}
    With these relations taken into account, the energy-momentum vector of
the tachyon in the $K'$ frame can be written as
    \begin{equation}    \label{rm09}
(p^{\prime\, i} )  = \left\{
\begin{array}{rl}
\displaystyle
\frac{m}{
\sqrt{ \frac{u^{\prime 2}}{c^2} - 1 }} \left( c, \mathbf{u^\prime} \right) , &
\displaystyle V < \frac{c^2}{ u}, \\[14pt]
\displaystyle
\pm \, m \, \left( 0,\, \mathbf{c} \right) , & \displaystyle V \to \frac{c^2}{ u} \mp 0, \\[14pt]
\displaystyle
\frac{- m}{
\sqrt{ \frac{u^{\prime 2}}{c^2} - 1 }} \left( c, \mathbf{u^\prime} \right) , &
\displaystyle V > \frac{c^2}{ u},
\end{array}
\right.
\end{equation}

    Under Lorentz transformations, the zeroth component of the energy-momentum
vector is transformed precisely as the time interval between events.
    Therefore, the sign of the energy of tachyons in the transition to
an inertial reference frame moving with a speed $V > c^2/u $ changes along
with the change in the direction of the time sequence of
events~\cite{Bilaniuk1962}.
    This coincidence led to the proposal~\cite{Bilaniuk1962} to use
the reinterpretation principle (following the ideas of
Stueckelberg~\cite{Stueckelberg1941} and Feynman~\cite{Feynman49}
to consider the positron as a negative-energy electron moving backwards
in time) to eliminate contradictions with causality violations: any object
with a negative energy that moves backwards in time must be reinterpreted
as its antiobject moving in the opposite direction in space, endowed with
positive energy, and traveling forward in time~\cite{Recami74}.
    A discussion of the relation between the reinterpretation principle
and the causality requirement can be found in~\cite{EinstainSb},
\cite{Grib74} and the references therein.

    A special situation regarding the assumption of the existence of tachyons
arises when observing them in a reference frame moving in the same direction
as the tachyon with the speed $V=c^2/u$.
    By Eqs.~(\ref{rm08}) and (\ref{rm09}), the tachyon must then have zero
energy and a nonzero momentum, and move at an infinite speed.

    Thus, in an inertial reference frame, the negative energies of ordinary
free particles of positive mass or tachyons are impossible if
the reinterpretation principle is taken into account.
    Negative energies can occur for particles with negative values of their
masses.
    Such particles were considered in theoretical physics already in
the 19th century (see book~\cite{JammerPM} and the references therein) and
are still being actively discussed~\cite{Bondi57}--\cite{NovikovKardashev11}.
%% \cite{Bondi57,Terletskii,LPPT,Bonnor89,NovikovKardashev11}.
    However, if we consider the least action principle, then only a positive
particle mass must be considered.
    For a negative mass, the action integral for a free particle would not have
a minimum value~\cite{LL_I}.

\vspace{4mm}
%%%% *****************************************************************
\section{\normalsize Energy in a noninertial reference frame in nonrelativistic
mechanics}

\hspace{\parindent}
    We show that in a noninertial reference frame, the energy of a particle
can take negative values in the absence of an external field.
   According to~\cite{LL_I} (see formula (39,6)),
the Lagrange function of a particle in an arbitrary noninertial reference
frame is given by
    \begin{equation}    \label{nm3}
L = \frac{mv^2}{2} + m \mathbf{v} [\mathbf{\Omega} \mathbf{r} ] +
\frac{m}{2} [\mathbf{\Omega} \mathbf{r} ]^2 - m \mathbf{W} \mathbf{r} - U,
\end{equation}
    where $\mathbf{r}$ is the particle radius vector,
    $\mathbf{W}$ is the acceleration of the translational motion of the
noninertial reference frame relative to an arbitrary inertial reference frame,
    $\mathbf{\Omega}$ is the angular velocity of rotation of the
noninertial frame, and $[\mathbf{\Omega} \mathbf{r}]$ denotes the vector
product of the corresponding vectors.

    Simple calculations (see~\cite{LL_I}, \S~39) show that
    \begin{equation}    \label{nm4}
\frac{\partial L}{\partial \mathbf{v}} = m \mathbf{v} +
m [\mathbf{\Omega} \mathbf{r} ].
\end{equation}
     Therefore, the energy of a point nonrelativistic particle in
a noninertial reference frame is equal to
    \begin{equation}    \label{nm5}
E = \mathbf{v} \frac{\partial L}{\partial \mathbf{v}} - L =
\frac{mv^2}{2} + U
- \frac{m}{2} [\mathbf{\Omega} \mathbf{r} ]^2 + m \mathbf{W} \mathbf{r} .
\end{equation}
    Obviously, for $\mathbf{\Omega}$ or $\mathbf{W}$ different from zero,
there always exist values of the coordinate $r$, at which the energy of
a free nonrelativistic particle is negative.

    We give some estimates. In the case without external field and for
a noninertial reference frame moving with a constant acceleration, we obtain
    \begin{equation}    \label{nm8}
E = \frac{mv^2}{2} + m \mathbf{W} \mathbf{r} < - mc^2  \ \
\Rightarrow \ \ |r| > \frac{c^2}{W}.
\end{equation}
    Thus, the negative energies greater than $mc^2$ in absolute value can be
observed only at large distances from the origin in the directions making
an obtuse angle with the acceleration of the reference frame:
$\mathbf{W} \mathbf{r} <0$.
    If an observer traveled from the origin with acceleration~$W$
until reaching such distances, then by nonrelativistic calculations,
the observer's speed must be equal to $\sqrt{2} c$.

    In the case of a uniformly rotating reference frame, the velocity of
the particle relative to the inertial frame whose origin coincides with
the origin of the rotating frame, is equal to
    \begin{equation}    \label{nm9}
\mathbf{v}_0 = \mathbf{v} + [\mathbf{\Omega} \mathbf{r}].
\end{equation}
     Energy~(\ref{nm5}) in the rotating frame can be expressed in terms of
the velocity~$\mathbf{v}_0$ as (see formula~(39,13) in~\cite{LL_I})
    \begin{equation}    \label{nm10}
E = \frac{mv_0^2}{2} + U - m [\mathbf{r}\mathbf{v}_0] \mathbf{\Omega}
= E_0 - \mathbf{M} \mathbf{\Omega},
\end{equation}
    where $E_0$ is the energy in the inertial frame and
$\mathbf{M}$ is the angular momentum of the particle (the same in the inertial
and rotating reference frames).
    In this case,
    \begin{equation}    \label{nm11}
E  < - mc^2 \ \ \Rightarrow \ \ r > \frac{c}{\Omega},
\end{equation}
    and negative energies greater than $mc^2$ in absolute value can be observed
only at distances from the origin at which the linear speed of rotation of
the coordinate axes is greater than the speed of light.
    We note that the motion of the coordinate axes, which are imaginary lines
of a coordinate system, with a speed greater than the speed of light does not
contradict the existence of a limit speed of signal transmission.
    The relative speeds of any physical objects in such a coordinate system
are equal to the speeds in the inertial frame and do not exceed the speed of
light.
    A relativistic analysis of a rotating coordinate system is carried out in
the next section.

    Thus, in a noninertial reference frame, there always exists a domain of
coordinate values where negative values of the energy of a free particle are
possible in the nonrelativistic case.
    The negative values comparable in absolute value with $m c^2$,
are attained in the domains where the characteristic speed of motion of
particles or coordinate axes is of the order of the speed of light.

    We further consider the cases of occurrence of negative energies in
noninertial coordinate systems in the relativistic case.

\vspace{4mm}
%%%% *****************************************************************
\section{\normalsize Energy of a point particle in a noninertial reference frame
in the relativistic case}
\label{secNewE}

{\renewcommand{\thefootnote}{\arabic{footnote}}

\hspace{\parindent}
    We assume that $x^i$ are coordinates related to an arbitrarily chosen
frame of reference and $g_{ik}$ is the metric tensor in these coordinates.
    In the absence of external fields, a particle moves along geodesic lines
in the space-time whose interval is equal to $ds^2 = g_{ik} dx^i dx^k$.
    The equations for these geodesics follow~\cite{Chandrasekhar}
from the Lagrangian
    \begin{equation}
L = \frac{g_{ik}}{2}\, \frac{ d x^i}{d \lambda} \frac{ d x^k}{d \lambda},
\label{Lgik}
\end{equation}
    where $\lambda$ is an affine parameter on a geodesic.
    For timelike geodesics,  $\lambda = \tau /m$,
where $\tau$ is the proper time of a moving particle of mass $m$.
    The generalized momenta are by definition equal to
    \begin{equation}
p_i  = \frac{\partial L}{\partial \dot{x}^i}
= g_{ik} \frac{d x^k}{d \lambda } , \ \ \ \
\dot{x}^i = \frac{d x^i}{d \lambda}.
\label{Lpdef}
\end{equation}
    The energy of a massive particle is
    \begin{equation}    \label{dE1}
E=p_0 c = m c\, g_{0k} \frac{d x^k}{d \tau }.
\end{equation}

    In the case of invariance under time shifts, i.e., if the metric
components $g_{ik}$ are independent of time, energy~(\ref{dE1}) is conserved
due to the Euler equations.
    In this case, the vector $(\zeta^i) = (1,0,0,0)$ is a timelike
Killing vector and the conserved energy~(\ref{dE1}) can be written as
    \begin{equation} \label{EKil}
E = m c^2\, \frac{dx^k}{ds}\,  g_{ik} \zeta^i =
m c^2 \, (u, \zeta) = c (p, \zeta).
\end{equation}
    We note that expression~(\ref{EKil}) is invariant under the choice of
the coordinate system but obviously depends on the specific choice of
the Killing vectors~$\zeta$.
    If there are two noncollinear timelike Killing vectors, then any linear
combination of such vectors is also a Killing vector and
energy values~(\ref{EKil}) corresponding to such Killing vectors
will be different.\footnote[1]{Translation of this phrase from
\href{https://doi.org/10.4213/tmf10453}{Russian original of the article}
in \href{https://doi.org/10.1134/S0040577923090088}{Theor. Math. Phys.} is incorrect.}
    If the Killing vector
$ ( \zeta^i ) = (1,0,0,0)$, is chosen in different coordinate systems, then
the energy values $E$ also are different in general, because passing to another
coordinate system changes not only the velocity values $u^i$, but also
the metric tensor.
    For a rotating frame of reference, we analyzed this question
in~\cite{GribPavlov2016d}, \cite{GribPavlov2019}.

}
%%%%%%%%%%%%%%%%%%%%%%%%%%%%%%%%%%%%%%%%%%%%%%%%%%%%%%%%%%%%%%%%%%%%%%

    If the vector $ ( \zeta^i ) $ is timelike, then the particle energy
in~(\ref{EKil}) is always positive (see problem~10.15 in~\cite{LPPT}).
    The particle energy becomes negative if the vector
$(\zeta^i) = (1,0,0,0)$ becomes spacelike in some domain.

    As an example, we consider a uniformly rotating frame of reference.
    Passing from the cylindrical coordinates $r'$, $\varphi'$, $z'$ of
the Minkowski space to rotating coordinates $r$, $\varphi$, $z$ by the formulas
    \begin{equation}    \label{v2}
r'=r, \ \ \ \ z'=z, \ \ \ \ \varphi' = \varphi - \Omega t,
\end{equation}
    we obtain the following expression for the interval:
    \begin{equation}    \label{v3}
d s^2 = (c^2 - \Omega^2 r^2)\, dt^2 + 2 \Omega r^2 d \varphi\, d t -
d r^{2} - r^{2} d \varphi^{2} - d z^{2}.
\end{equation}
    On the surface $r=c/\Omega$, the metric component $g_{00}$ vanishes,
but the metric is nondegenerate in this case:
$ {\rm det}\! \left( g_{ik} \right) = - r^2$.
    The requirement that the interval of the worldline of a particle be
nonnegative implies that outside the surface $r=c/\Omega$,
no physical body can be at rest in a given coordinate system.
    In the course of time, any body must move in the direction of
increasing~$\varphi$, i.e., in the direction opposite to the direction of
rotation of the coordinate system.
    The occurrence of a static limit is similar to the case of a rotating black
hole.
    Here, a three-dimensional ``rigid grid'' corresponding to the rotating
coordinates ``cannot be made of material bodies
(`welded' of rods)''~\cite{NovikovFrolov}, \S~4.2.
    Beyond the static limit, such a grid would move with respect to any
 observer with a speed greater than the speed of light.

    We note that before the appearance of the exact solution of the Einstein
equations describing the metric of a rotating black hole, i.e., the Kerr
solution~\cite{Kerr63}, the impossibility of realizing the coordinate
system by fixed rigid bodies led some authors to conclude that such
a coordinate system could not be used (see the discussion and the references
in~\cite{GribPavlov2019}).
    However, the wide use of the Kerr solution, which has a static limit
(outside the black hole event horizon), shows that the impossibility of
constructing a rigid coordinate grid is an acceptable feature of
the coordinate system.
    We also note that the implementation of a rotating frame of reference
by (absolutely) rigid bodies is fundamentally impossible when relativistic
effects are taken into account not only beyond the static limit but even
near the origin due to the failure of the concept of an absolutely rigid
body under rotation, as was shown in~\cite{Erenfest1909}.

    Energy~(\ref{EKil}) in a uniformly rotating coordinate system is equal
to~\cite{GribPavlov2019}
    \begin{equation}    \label{v10v}
E = E' + \Omega L'_z,
\end{equation}
    where $ E' $ and $ L'_z $ is the energy and the projection of the angular
momentum on the axis of rotation $(OZ')$ in an inertial system with
coordinates $r'$, $\varphi'$, $z'$.
    We note that this expression formally coincides with nonrelativistic
expression~(\ref{nm10}) if we set $\mathbf{\Omega} = (0, 0, - \Omega)$,
but the energy $ E' $ and the angular momentum projection $ L'_z $
in~(\ref{v10v}) are defined by relativistic formulas.

    For the Killing vector of translations along the coordinate~$x^0=c t$
in a rotating frame of reference, we have
    \begin{equation}    \label{Ki2}
(\zeta , \zeta)= 1 - \frac{\Omega^2 r^2}{c^2} .
\end{equation}
     Thus, beyond the static limit, the vector $\zeta$ becomes spacelike,
and the particle energy~$E$ beyond the static limit in the rotating system can
take negative values that are arbitrarily large in absolute value (for details,
see~\cite{GribPavlov2019}).

    The presence of negative energy values~(\ref{EKil}) for a spacelike
vector $(\zeta^i)$ does not violate the energy dominance condition
(see \S~8.3 in~\cite{GMM}).
    To prove this, we note that expression~(\ref{EKil}) can be obtained by
integrating the energy-momentum tensor of a point particle with
action~(\ref{rm01}) (see \S 12.2 in~\cite{Weinberg}) over the
hypersurface $\Sigma$ perpendicular to $\zeta$:
    \begin{equation} \label{NEnergy2}
E = \int_\Sigma T_{ik}\, \zeta^i \, d \sigma^k ,
\end{equation}
    \begin{equation} \label{NEnergy5}
T^{ik}(x) = - \frac{2 c}{\sqrt{ |g|}} \, \frac{\delta S}{\delta g_{ik}}
= \frac{m c^2 }{\sqrt{|g|}} \int \! ds \, \frac{d x_p^i}{d s}\,
\frac{d x_p^k}{d s}\, \delta^4 ( x - x_{p}) ,
\end{equation}
    where $x_p$ is the world point at which the particle is localized and
$g= {\rm det}(g_{ik})$.
    The expression for $T^{ik}(x) \sqrt{ |g|} $ is independent of the choice
of the reference frame; in particular, it looks the same as in the inertial
frame of reference, which proves the energy dominance of the energy-momentum
tensor of a positive-mass point particle.
    This proof of energy dominance holds for an arbitrary curved space-time,
and hence the presence of negative-energy states for ordinary point particles
cannot be a source of a wormhole metric.

    We note that expression~(\ref{dE1}) also determines the particle energy
in the case where there is no invariance in time and energy is not conserved.

\vspace{4mm}
%%%% *****************************************************************
\section{\normalsize Milne universe}

\hspace{\parindent}
    We consider the space-time with the interval
    \begin{equation}
d s^2 = c^2 d t^2 - c^2 t^2 \left( d \chi^2 + \sinh^2 \chi d \Omega^2 \right),
\label{Le6}
\end{equation}
    where $d \Omega^2 = d \theta^2 + \sin^2 \theta \, d \varphi^2 $
and the coordinate $\chi $ ranges $[0, + \infty) $.
    This space-time, called the Milne universe~\cite{Milne},
is a part of a flat space-time, because the change of coordinates
    \begin{equation}
T = t \cosh \chi, \ \ \ \ r = c t \sinh \chi , \ \ \ \ cT > r > 0
\label{Le7}
\end{equation}
    takes~(\ref{Le6}) to the form of a Minkowski space interval in spherical
coordinates,
    \begin{equation}
ds^2 = c^2 d T^2 - d r^2 - r^2\, d \Omega^2.
\label{Le8}
\end{equation}

    The reference frame corresponding to the coordinates $t$ and $\chi$
can be realized as follows~\cite{ZeldovichNovikovSEU}.
    In the flat Euclidean space at the initial time $T=0$,
a set of particles moving with all possible velocities $u<c$ is emitted
from the origin.
    We neglect the back reaction of the energy density of the emitted particles
on the space-time metric, assuming that it remains Euclidean.
    The proper time $t$ of a pais related to the coordinate time $T$ as
    \begin{equation}
t = T \sqrt{1 - \frac{u^2}{c^2} }.
\label{SV1}
\end{equation}
    and hence the Euclidean coordinates and the time of a particle moving
with a speed $u$ are equal to
    \begin{equation}
T = \frac{t}{ \sqrt{1 - \frac{u^2}{c^2} } }, \ \ \ \ r= u T =
t \frac{u}{ \sqrt{1 - \frac{u^2}{c^2} } } .
\label{SV2}
\end{equation}
    Setting $ \cosh \chi = 1/ \sqrt{\mathstrut 1 - u^2/c^2 }$, we obtain
formulas~(\ref{Le7}) for the transition to the Milne coordinates.

    The distance from the origin to the point with a prescribed value of~$\chi$
in metric~(\ref{Le6}) is equal to $D = c t \chi$.
    Using this value for the radial coordinate in~\cite{Ellis93},
we write interval~(\ref{Le6}) as
    \begin{eqnarray}
ds^2 = \left( 1- \frac{D^2}{c^2 t^2} \right) c^2 d t^2 + 2 \frac{D}{t} dt d D -
d D^2 - c^2 t^2 \sinh^2 \! \left( \frac{D}{ct} \right) d \Omega^2.
\label{Le9}
\end{eqnarray}
    The presence of the off-diagonal term in metric~(\ref{Le9}) and
the requirement that $ds^2 \ge 0$ imply that for~$D > D_s = ct $,
not a single physical object can be at rest in the coordinates $ D, \theta, \phi$.
    The value of $D_s$ corresponding to $\chi = 1$ plays the role of
the static limit for a rotating black hole in the Boyer-Lindquist
coordinates~\cite{BoyerLindquist67}.

    The energy of a particle of mass~$m$ in the reference frame with
coordinates $(t, D, \theta, \phi)$, Eq.~(\ref{dE1}), is
    \begin{equation}
E = c g_{0k} \frac{d x^k}{d \lambda}
= m c^2 \frac{d t}{ d \tau} \left(
1 - \frac{D^2}{c^2 t^2} + \frac{D}{c^2 t} \frac{d D}{d t} \right)
= m c^2 \frac{d t}{ d \tau} \left( 1 + \chi t \frac{ d \chi}{d t} \right).
\label{Le10}
\end{equation}
    Because the components of metric~(\ref{Le9}) depend on time, the particle
energy is not conserved on geodesics in this reference frame.

\begin{figure}[ht]
\centering
\includegraphics[width=59mm]{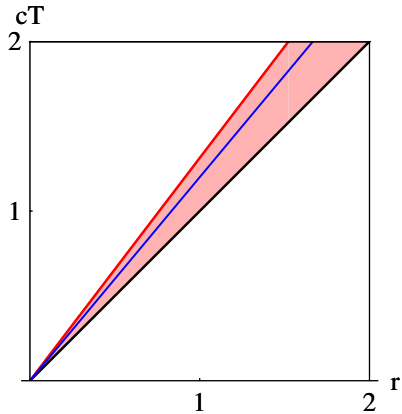}
\caption{Domain of flat coordinates $(T,r)$ where negative particle energies
are possible in the reference frame $(t, D)$.}
\label{FigM1}
\end{figure}

    From~(\ref{Le10}), we obtain that the particle energy is negative if
    \begin{equation}
E < 0 \ \ \Leftrightarrow \ \ \chi t \frac{ d \chi}{d t} < - 1.
\label{Le12}
\end{equation}
    The particle energy in the reference frame $(t, D, \theta, \phi)$
is equal to zero if
    \begin{equation}    \label{Le13a}
E = 0 \ \ \Leftrightarrow \ \ \frac{ d \chi}{d t} = -\frac{1}{\chi t}.
\end{equation}
    It follows from the condition $ds^2 \ge 0$ for~(\ref{Le6}) that
    \begin{equation}
t \left| \frac{ d \chi}{d t} \right| \le 1 ,
\label{Le11}
\end{equation}
    and therefore negative particle energies are possible for $\chi >1$,
and the zero energy, for $\chi \ge 1$, i.e., beyond the static limit.
    In the coordinates $T,r$ by~(\ref{Le7}),
this domain is defined by the inequalities $1 \le cT/r \le \coth 1 \approx 1.313$
(see Fig.~\ref{FigM1}).
%%%%%%%%%%%%%%%%%%%%%%%%%%%%%%%%%%%%%%%%%%%%%%%%%%

    Formula~(\ref{Le7}) implies a relation between the speeds in
the coordinates $t, \chi$ and $T,r$:
    \begin{equation}
t \frac{d \chi}{d t} = \frac{ \displaystyle
\frac{dr}{d T} - c \tanh \chi}{ \displaystyle c - \frac{d r }{d T} \tanh \chi}.
\label{Le14}
\end{equation}
    Therefore, by~(\ref{Le12}), a particle at rest in the inertial reference
frame $T,r$ has zero energy in the reference frame $t, \chi$
if $ \chi \tanh \chi = 1$,  i.e., at ($\chi \approx 1.1997$)
(the blue line in Fig.~\ref{FigM1}) and has a negative energy for
$ \chi \tanh \chi > 1 $
(the shaded area in Fig.~\ref{FigM1} below the blue line).

    Thus, with a certain choice of a noninertial frame of reference,
the energy value measured in it can be zero and negative even for particles
at rest in some inertial frame of reference.

    If a body in the domain $D > ct$ beyond the static limit splits into two
fragments, one with a negative energy and the other with a positive energy,
then the energy of the second fragment is greater than the energy of
the original body, as in the well-known Penrose process.
    Can an observer located nearby the origin get a gain in energy?
    We note that at small distances from the origin (for $D/(ct) =\chi \ll 1$),
metric~(\ref{Le9}) corresponds to the spherical-coordinate metric of
the Minkowski space
    \begin{equation}
ds^2 = c^2 d t^2 - d D^2  -D^2 d \Omega^2,
\label{Le9mD}
\end{equation}
    and hence energy~(\ref{Le10}) is equal to the energy in the inertial system
of the flat space-time
    \begin{equation}
E_u = m c^2 \frac{d t}{ d \tau} \approx m c^2 \frac{d T}{ d \tau},
\label{Le10mD}
\end{equation}
    because it follows from~(\ref{Le7}) that $t \approx T$ for $\chi \ll 1$.

    In the case of a decay in accordance with the Penrose process, the first
fragment with negative~(\ref{Le10}) has a negative velocity $d\chi/ d t$.
    If the positive-energy fragment also moves towards the origin,
then, by~(\ref{Le10}), the absolute value of its velocity $|d\chi/ d t|$ is
less than that of the negative-energy fragment.
    By~(\ref{Le14}), the same holds for velocities $d r / d T$,
if the fragments move towards the origin.
    Thus, for an observer near the origin of the $t, D$ coordinates,
the negative-energy fragment that from the decay of the original body arrives
sooner, and by the time of its arrival, its energy~(\ref{Le10})
(which is not conserved) becomes positive.
    The observer is unable to gain energy.

    In the case of the Kerr metric, the energy is a conserved quantity.
    An observer at a large distance from a rotating black hole can observe
an energy gain if the positive-energy fragment resulting from the decay of
a particle by the Penrose process in the ergosphere recedes to a large distance
from the black hole.
    The energy is also conserved in the case of a rotating frame of reference,
which allows a process similar to the Penrose process in a rotating coordinate
system~\cite{GribPavlov2019}.

    We note that the presence of negative-energy states in the Milne universe
leads to the transition to a vacuum state different from the vacuum in
the Minkowski space.
    This can be interpreted as creation of quasiparticles
(see~\S~5.3 in~\cite{BirDev}).
    But in this case, these are virtual particles (see~\S~9.8 in~\cite{GMM}),
which appear in the excitation of the corresponding noninertial particle
detector and do not exert back reaction on the space-time metric.

\vspace{4mm}
%%%% *****************************************************************
\section{\normalsize Rindler space-time}

\hspace{\parindent}
    The widely known Unruh effect~\cite{Unruh76} consists in the fact
that a detector moving with a constant acceleration a must detect the presence
of quasiparticles whose spectrum exhibits the thermal character corresponding
to temperature
    \begin{equation}
T_a = \frac{\hbar a}{2 \pi c k_B} \approx 4.055 \cdot 10^{-21} K \frac{a}{
1\, {\rm m/s}^2},
\label{Un1}
\end{equation}
    where $\hbar$ is the reduced Planck constant and $k_B$ is
the Boltzmann constant.

    To describe the motion with a constant acceleration, it is convenient
to use the Rindler coordinates~\cite{Rindler1966} which are related to
the usual Cartesian coordinates $(ct, x^1,x^2,x^3)$ as
    \begin{equation}
T = \frac{c}{2a} \ln \left| \frac{ct +x^1}{ct- x^1} \right|, \ \ \ \
\rho = \frac{a}{c^2} \left( (x^1)^2 - (ct)^2 \right).
\label{Un2}
\end{equation}
    In these coordinates, the metric becomes
    \begin{equation}
d s^2 = \rho a \, d T^2 - \frac{c^2}{4 \rho a} \, d \rho^2 - d (x^2)^2
- d (x^3)^2.
\label{Un3}
\end{equation}
    The space-time splits into four domains $\mathbf{R}$,
$\mathbf{L}$, $\mathbf{F}$, $\mathbf{P}$ bounded by the isotropic hypersurfaces
$\rho = 0$, $ \tau = - \infty $ and $\rho = 0$, $\tau = + \infty $:
    \begin{eqnarray}
\mathbf{R}: & \ \  x^1 \ge 0,& \ \ \ \ \ x^1 > c |t|, \nonumber \\
\mathbf{L}: & \ \  x^1 < 0,& \ \ \ \  |x^1| > c |t|, \nonumber \\
\mathbf{F}: & \ \ \ t \ge 0,& \ \ \ \  |x^1| < c t, \nonumber \\
\mathbf{P}: & \ \ \ t \le 0,& \ \ \ \  |x^1| < c |t|. \nonumber
%% \label{URLFP}
\end{eqnarray}

    In the domain $\mathbf{R}$, the coordinate $T$ is the time, and $\rho$
is the spatial coordinate.
    The transformations inverse to~(\ref{Un2}) have the form
    \begin{equation}
t = \sqrt{\frac{\rho}{a}} \sinh \left( \frac{a T}{c} \right), \ \ \ \
x^1 = c \sqrt{\frac{\rho}{a}} \cosh \left( \frac{a T}{c} \right).
\label{Un4}
\end{equation}
    Therefore, the hyperbolas $\rho =\rho_0 = {\rm const}$,
$({x^2, x^3}) = {\rm const}$ are the worldlines of particles moving with
a constant proper acceleration $c \sqrt{a/\rho_0}$ along the $x^1$ axis
and $ T \sqrt{\rho_0 a} /c $ is the proper time of these particles.
    The reference frame formed by a set of such particles (``observers'')
covers the entire domain $\mathbf{R}$.
    For the observers from $\mathbf{R}$, the surface $\rho = 0$, $ T = + \infty $
is the future event horizon, and the surface $\rho = 0$, $ T = - \infty $
is the past event horizon (for details, see~\S~12.1 in~\cite{GMM}).

    The vector $\zeta = (1,0,0,0)$ is a Killing vector for metric~(\ref{Un3}),
which is timelike in the domain $\mathbf{R}$.
    The particle energy calculated by formulas~(\ref{dE1}), (\ref{EKil})
is equal to
    \begin{equation}
E = ma \frac{d t}{d \tau}(x^1 - v^1 t) = \frac{ mc^2 }{\sqrt{ 1 - \frac{v^2}{c^2} } }
\, \frac{\rho}{(x^1 + v^1 t)} ,
\label{Un5}
\end{equation}
    where $v^1 = d x^1 / (d t)$, $ v $ is the velocity in the Cartesian
coordinates $ (ct, x^1, x^2, x^3) $.
    It follows from~(\ref{Un5}) that the energies can be only positive in
the domain $\mathbf{R}$ and only negative in the domain $\mathbf{L}$,
while in the domains $\mathbf{F}$ and $\mathbf{P}$ the energies can be of
opposite signs and equal to zero for $v^1 = x^1/t$
(because the inequality $|x^1 /t| < c$ holds in the domains $\mathbf{F}$
and $\mathbf{P}$).
    The pattern of possible energy signs~(\ref{Un5}) in the Rindler space-time
is similar to the situation with energy signs for a nonrotating black holes in
Kruskal-Szekeres coordinates (see Fig.~1 in~\cite{GribPavlov2010NE}).

\vspace{4mm}
%%%% *****************************************************************
\section{\normalsize Conclusions}
\label{secEPenr}

\hspace{\parindent}
    In this paper, we considered various cases of the existence of particles
with negative and zero energy in the absence of external force fields.
    When measuring the energy in an inertial frame of reference, the energy of
ordinary massive particles is always positive. In the case of hypothetical
particles moving faster than light, tachyons, the energy is also positive
in any inertial frame of reference under the assumption that the principle of
reinterpretation is valid.
    The negative energy of a free particle in an inertial frame of reference
is possible only in the hypothetical case of negative particle masses.

    We showed that in a noninertial reference frame, there always exists
a domain of coordinates where the energy of ordinary positive-mass particles
is negative or zero in the nonrelativistic case.

    We considered the general definition of energy in noninertial reference
frames in the relativistic case and showed that the definitions of energy of
a relativistic particle in terms of the zeroth component of canonical
momentum~(\ref{dE1}) in terms of the corresponding Killing vector~(\ref{EKil})
(if it exists) and in terms of the energy-momentum tensor of a point
particle~(\ref{NEnergy5}) lead to the same results.
    If the vector of time translations is spacelike in a certain domain, then
the particle energy in this domain can be negative or zero.
    However, despite the existence of negative-energy states, the requirement
of energy dominance is not violated.
    Thus, the usual particles whose energy becomes negative for some
noninertial observer cannot be sources of wormhole-like metrics in
the general theory of relativity.

    We considered two examples of a flat space-time in noninertial coordinates:
the Milne universe and the Rindler space.
    For these spaces, we found domains where states with negative and
zero energy occur.
    Using the Milne universe as an example, we showed that in a noninertial
frame of reference, the corresponding relativistic energy can be zero and
negative even for particles at rest in some inertial frame of reference.

\vspace{4mm}
%%%% ****************************************************************
{\bf Conflicts of interest.}
The authors declare no conflicts of interest.

\vspace{4mm}
%%%% ****************************************************************
{\bf Acknowledgments.}\,
    The authors thank the participants of the MQFT-2022 conference for
the discussion of the talk, and D.\,I. Kazakov for formulating the question
that underlies the title of the paper.

    The work was supported by the Russian Science Foundation
(grant No. 22-22-00112).

\vspace{2mm}
%%%% ****************************************************************

\end{document}